\documentclass[10pt,preprint]{emulateapj}
\def\msun{\rm M_{\sun}}

\def\lsun{\rm L_{\sun}}
\def\micron{$\mu$m}
\def\microns{$\mu$m}
\def\mdot{\rm \dot{M}}

\begin{document}

\slugcomment{{\sc Accepted to ApJ:} June 18, 2007} 

\shortauthors{Espaillat et al.}
\shorttitle{The Transitional Disk of CS Cha}

\title{Probing the Dust and Gas in the Transitional Disk of CS Cha with $Spitzer$}

\author{Catherine Espaillat\altaffilmark{1}, Nuria Calvet\altaffilmark{1}, Paola D'Alessio\altaffilmark{2}, Edwin Bergin\altaffilmark{1}, Lee Hartmann\altaffilmark{1}, \\Dan Watson\altaffilmark{3}, Elise Furlan\altaffilmark{4,5}, Joan Najita\altaffilmark{6}, William Forrest\altaffilmark{3}, Melissa McClure\altaffilmark{3}, \\Ben Sargent\altaffilmark{3}, Chris Bohac \altaffilmark{3}, \& Samuel T. Harrold\altaffilmark{3}}

\altaffiltext{1}{Department of Astronomy, University of Michigan, 830 Dennison Building, 500 Church Street, Ann Arbor, MI 48109, USA}
\altaffiltext{2}{Centro de Radioastronom\'{i}a y Astrof\'{i}sica, Universidad Nacional Aut\'{o}noma de M\'{e}xico, 58089 Morelia, Michoac\'{a}n, M\'{e}xico}
\altaffiltext{3}{Department of Physics and Astronomy, University of Rochester, NY 14627-0171, USA}
\altaffiltext{4}{NASA Astrobiology Institute, and Department of Physics and Astronomy, UCLA, Los Angeles, CA 90095, USA}
\altaffiltext{5}{NASA Postdoctoral Fellow}
\altaffiltext{6}{NOAO, Tucson, AZ 85719, USA}

\email{ccespa@umich.edu}

\begin{abstract}
Here we present the $Spitzer$ IRS spectrum of CS Cha, a member of the $\sim$2 Myr old Chamaeleon star-forming region, which reveals an optically thick circumstellar disk truncated at $\sim$43 AU, the largest hole modeled in a transitional disk to date.  Within this inner hole, $\sim$5$\times$10$^{-5}$ lunar masses of dust are located in a small optically thin inner region which extends from 0.1 to 1 AU.  In addition, the disk of CS Cha has bigger grain sizes and more settling than the previously modeled transitional disks DM Tau, GM Aur, and CoKu Tau$/$4, suggesting that CS Cha is in a more advanced state of dust evolution.  The $Spitzer$ IRS spectrum also shows [Ne II] 12.81 {\micron} fine-structure emission with a luminosity of 1.3 $\times$ 10$^{29}$ ergs s$^{-1}$, indicating that optically thin gas is present in this $\sim$43 AU hole, in agreement with H$_{\alpha}$ measurements and a UV excess which indicate that CS Cha is still accreting 1.2 $\times$ 10$^{-8}$ $\msun$ yr$^{-1}$.  We do not find a correlation of the [Ne II] flux with L$_{X}$, however, there is a possible correlation with ${\mdot}$, which if confirmed would suggest that EUV fluxes due to accretion are the main agent for formation of the [Ne II] line.  
\end{abstract}


\section{Introduction}
\label{sec:int}


The
$Spitzer$ $Space$ $Telescope$ \citep{werner04} has dramatically improved the ability to study the dust in disks by giving us detailed spectral energy distributions (SEDs) in the mid-infrared, where the dust dominates the emission.  $Spitzer$ Infrared Spectrograph 
    \citep[IRS;][]{houck04} observations of T Tauri stars have provided strong 
    evidence of dust evolution, particularly in the well surveyed $\sim$1-2 Myr old Taurus-Auriga star-forming region \citep{furlan06}.  The most notable evidence of dust evolution lies in the observations of transitional
 disks.  These disks have characteristics that fall 
 between those objects that have clear evidence for disks and those objects 
 with no disk material; the SEDs point to radial gaps in the disk.  
  All transitional disks have a significant
  deficit of flux in the near-IR which has been
   explained by modeling the disks with truncated optically thick disks \citep{uchida04, calvet05, dalessio05}.  
    
CoKu Tau$/$4 is one of these transitional disks in Taurus \citep{forrest04, dalessio05}.  The SED of this object is photospheric below 8 {\micron} and rises at longer wavelengths, showing weak silicate emission at 10 {\micron}; this can be explained by the emission of the frontally illuminated edge or ``wall" of an outer disk truncated at 10 AU from the star \citep{dalessio05} and the weak silicate emission arises in the atmosphere of the wall.  There are no small grains in the inner disk, in agreement with CoKu Tau$/$4's classification as a non-accreting, weak T Tauri star.  Hydrodynamical simulations indicate that this gap may be due to the formation of a planet \citep{quillen04}.  Other studies point to photoevaporation of the disk by the star \citep{alexander06}.
Similar to CoKu Tau$/$4, DM Tau has been modeled with an inner disk region free of small dust and a truncated optically thick disk at 3 AU \citep[C05]{calvet05}.  However, this star is still accreting, indicating that gas should still remain in the inner disk region.  GM Aur, which is also accreting, was modeled by C05 with an optically thick disk truncated at 24 AU and a small amount of dust in the inner disk which leads to its observed near-IR excess.
This disk gap could also be due to a planet \citep{rice03}.  TW Hya, a transitional disk in the $\sim$10 Myr old TW Hya association \citep{jayawardhana99, webb99}, has an outer optically thick disk with a 4 AU hole filled with an inner optically thin region populated by a small amount of dust \citep{calvet02, uchida04}.  

Not only has
$Spitzer$ been a powerful tool in exploring the dust evolution in circumstellar disks, it also has the potential to access new probes of the gas evolution in these disks, an area of study which is still in its infancy and largely uncertain.  
High-resolution molecular spectroscopy has permitted us to access abundant molecular tracers within the inner disk such as H$_2$ and CO \citep{bergin04, najita07}.  
A new gas diagnostic has emerged in the mid-infrared with the detection of [Ne II] emission with $Spitzer$.  [Ne II] emission has been detected from a few T Tauri stars that were 
observed as part of the FEPS Spitzer Legacy program 
\citep[P07]{pascucci07} as well as in the transitional disks CS Cha, DM Tau, and TW Hya (this paper).  


We will present and analyze Spitzer IRS data of CS Cha, a transitional disk in the $~\sim$2 Myr old Chamaeleon star-forming region \citep{luhman04}.

\section{Observations}
\label{sec:obs}

CS Cha was observed by the $Spitzer$ IRS instrument on 11 July 2005 in Staring Mode (AOR ID: 12695808).
We used the short- and long-wavelength, low-resolution (SL, LL) modules of IRS 
at a resolving power of $\lambda$/$\delta$$\lambda$ = 60-100.  
The total exposure time was 28 and 12 seconds for SL and LL respectively.  We reduced the data with the SMART package \citep{higdon04}.  We 
follow the same reduction procedure as C05.  
The 12.81 {\micron} [Ne II] line is not extended and is consistent with the CS Cha point source.

The photometry at 24 and 70 {\microns} was extracted from the MIPS
post-bcd mosaicked images (AOR ID: 3962112) using the aperture photometry routine in the
Astronomical Point Source Extraction (APEX) software package developed
by the SSC \citep{makovoz05}.  An aperture of radius 14.94
$^\prime$$^\prime$, with a sky annulus between 29.88$^\prime$$^\prime$ and 42.33$^\prime$$^\prime$, was used at
24 {\microns} while an aperture of radius 29.55$^\prime$$^\prime$, with a sky
annulus between 39.40$^\prime$$^\prime$ and 68.95$^\prime$$^\prime$, was used at 70 {\microns}.  Both fluxes
were aperture corrected by factors of 1.143 at 24 {\microns} and 1.298 at
70 {\microns}, as calculated by \citet{su06}.  No color correction was
applied.

\section{Analysis}
\label{sec:res}
\subsection{Dust Properties}

\begin{deluxetable}{l l}
\tabletypesize{\scriptsize}
\tablewidth{0pt}
\tablecaption{Stellar and Model Properties\label{tab:prop}}
\tablehead{
\colhead{Stellar Parameters} & \colhead{}}
\startdata
$M_{*}$ $(M_{\sun})$................. & 0.91  \\
$R_{*}$ $(R_{\sun})$.................. & 2.3  \\
$T_{*}$ $(K)$......................... & 4205 \\
$L_{*}$ $(\lsun)$....................... & 1.5 \\
$\mdot$ $(M_{\sun} yr^{-1})$............... & 1.2 $\times$ 10$^{-8}$ \\
Distance (pc)................ & 160\\
$A_{V}$.............................. & 0.8 \\
\hline
Wall Parameters\\
\hline
$R_{wall}$ $(AU)$.................. & 42.7  \\
$a_{min}$ ({\micron}).................. & 0.005  \\
$a_{max}$ ({\micron}).................. & 5  \\
$T_{wall}$ $(K)$......................... & 90 \\
$z_{wall}$ $(AU)^{1}$............................. & 5 \\
\hline
Optically Thick Outer Disk Parameters\\
\hline
$R_{d,out}$ $(AU)$.................. & 300 \\
$\epsilon$......................... & 0.01 \\
$\alpha$.................. & 0.005 \\
$M_{d}$ $(M_{\sun})$................ & 0.04\\
\hline
Optically Thin Inner Region Parameters\\
\hline
$R_{in, thin}$ $(AU)$.................. & 0.1 \\
$R_{out, thin}$ $(AU)$.................. & 1 \\
$a_{min, thin}$ ({\micron}).................. & 1.9  \\
$a_{max, thin}$ ({\micron}).................. & 2.1  \\
$M_{d,thin}$ $(M_{\sun})$................ & 1.7 $\times$ $10^{-12}$
\enddata
\tablenotetext{1}{z$_{wall}$ is the height above the disk midplane}

\end{deluxetable}

\begin{figure}
\epsscale{0.7}
\plotone{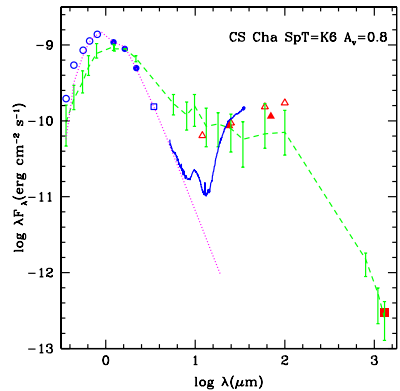}
\caption{SED of CS Cha.  Optical (open circles), J,H,K (filled circles), L-band (open square), $Spitzer$ IRS data (blue solid line), IRAS (open triangles), MIPS (closed triangles) and sub-mm (closed square) are shown.  Red corresponds to observed magnitudes and blue symbols are dereddened magnitudes.  The short-dashed green line with quartiles is the median SED of Taurus and the dotted magenta line is the stellar photosphere.  [See the electronic edition of the Journal for a color version of this figure.]
}
\label{figsed}
\end{figure}

Figure ~\ref{figsed} shows the SED of CS Cha consisting of optical \citep{gauvin92}, 2MASS, L-band \citep{luhman04}, Spitzer IRS, IRAS \citep{gauvin92}, MIPS, and millimeter \citep{henning93} data.  We also show the median SED of Taurus \citep{dalessio99, furlan06} which emphasizes the stark deficit of flux in the near infrared, an indicator of an inner disk hole, which is characteristic of transitional disks.  

A distance of 160 pc to Chamaeleon I \citep{whittet97} and a spectral type of K6 from \citet{luhman04} are adopted.  The extinction is calculated by fitting the observed photospheric colors to the photosphere calibrated by a standard K6 star with an effective temperature of 4205 K \citep{KH95}.  Data are then dereddened with an A$_{V}$ of 0.8 and the \citet{mathis90} reddening law.  Stellar parameters (M$_{*}$, R$_{*}$, L$_{*}$) are listed in Table 1; the mass was derived from the Baraffe evolutionary tracks \citep{baraffe02}.  An inclination angle of 60 degrees to the line of sight is adopted.  

The SED suggests a UV excess which is thought to be formed in the accretion shock on the stellar surface \citep{calvetgullbring98}.  This is supported by H$_{\alpha}$ equivalent widths of 20 {\AA} \citep{luhman04} and 65 {\AA} \citep{hartigan93}, which indicate CS Cha is accreting \citep{whitebasri03}. We calculate a mass accretion rate of 1.2 $\times$ 10$^{-8}$ $\msun$ yr$^{-1}$ from this U-band excess following \citet{gullbring98}, with a typical uncertainty of a factor of 2 or 3 \citep{calvet04}.  


\begin{figure}
\epsscale{0.7}
\plotone{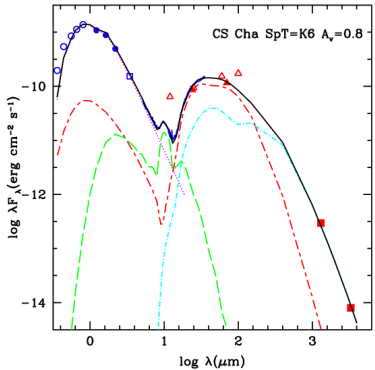}
\caption{SED and model of CS Cha.  The solid black line is the best fit model with parameters from Table 1.  Separate model components are also shown: wall (red short-long-dash line), optically thick disk (blue dot-dash), optically thin inner region (green long-dashed), stellar photosphere (magenta dotted line).  Note that the model does not fit the IRAS points.  IRAS has a larger field of view than $Spitzer$ which could overreport the flux, especially at 100 {\microns}.  However, the model agrees with the MIPS data, which has a smaller FOV than IRAS.  We also include SEST 3.3 mm data \citep{lommen07}. [See the electronic edition of the Journal for a color version of this figure.]
}
\label{figmodel}
\end{figure}

We model CS Cha as a truncated optically thick disk with a frontally illuminated wall and an inner optically thin region.  The black solid line in Figure ~\ref{figmodel} is the best fit model to the observations; different model components are represented by the broken lines.   We use a grain-size distribution that follows a power-law of the form a$^{-3.5}$, where $a$ is the grain radius.   
The structure and emission of the optically thick disk (Figure ~\ref{figmodel}) is calculated with models including dust settling following \citet{dalessio06}, where we use millimeter fluxes \citep{henning93, lommen07} and MIPS to constrain the outer disk properties. Input parameters are the stellar properties and the mass accretion rate of the disk ($\mdot$), the viscosity parameter ($\alpha$), the dust composition, and the settling parameter $\epsilon=\zeta_{up}/\zeta_{st}$, i.e. the mass fraction of the small grains in the upper layers relative to the standard dust-to-gas mass ratio \citep{dalessio06}.  Table 1 shows all relevant values.  The disk mass, 0.04 M$_{\sun}$, is determined by the mass surface density which is proportional to $\mdot/\alpha$.  
This result is not inconsistent with the disk mass of 0.021 M$_{\sun}$ derived by \citet{henning93} from 1.3 mm fluxes.  
 
The optically thick disk has an edge or "wall" directly exposed to stellar radiation (Figure ~\ref{figmodel}).  The radiative transfer in the wall atmosphere is calculated following \citet{dalessio05} with the following inputs: stellar mass (M$_{*}$), stellar radius (R$_{*}$), $\mdot$, stellar effective temperature (T$_{*}$), maximum and minimum grains sizes, temperature of the optically thin wall atmosphere (T$_{wall}$).  See Table 1 for the wall's location (R$_{wall}$), maximum grain size (a$_{max}$), and other parameters.       

The emission at 10 {\micron} comes from the optically thin inner region (C05) which extends up to 1 AU from the star and has a minimum grain size of 1.9 {\micron} and a maximum grain size of 2.1 {\micron} (Figure ~\ref{figmodel}).  
The total emission is scaled to the vertical optical depth at 10 {\micron}, $\tau_{0} \sim 0.009$.  This region is populated by $\sim$ 10$^{-12}$ M$_{\sun}$ or $\sim$5$\times$10$^{-5}$ M$_{moon}$ of uniformly distributed dust.  This small amount of dust leads to the slight 5 -- 8 {\micron} excess above the photosphere (most easily seen in Fig. 3).  The dust in this region is composed of 88\% amorphous silicates, 5\% of amorphous carbon, 5\% organics, 1\% troilite and less than 1\% of enstatite and forsterite. 

Figure ~\ref{figvs} illustrates the necessity of larger grains in both the wall and optically thin inner region to fit the observations.  The solid line corresponds to the model
  of the wall and optically thin region with the larger grain sizes
  which we adopted above. The dotted line corresponds to models with
  the smaller, ISM-like grains, as are found in the Taurus transitional disks CoKu Tau$/$4, DM Tau, and GM Aur.  Figure ~\ref{figvs} shows that larger grains have better agreement with the slope of the SED beyond 15 {\microns} and are necessary to produce the slight near-infrared excess, the shape of the 10 {\micron} feature, and the $>$20 {\micron} IRS emission in CS Cha.

\subsection{Gas Properties}

\begin{figure}
\epsscale{0.7}
\plotone{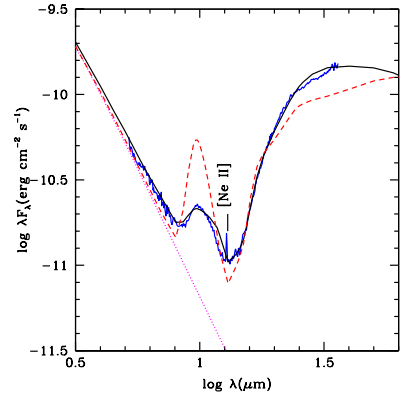}
\caption{Models of CS Cha with big grains versus small grains.  The solid black line corresponds to the model that uses big grains and it is the best-fit disk model shown in Figure~\ref{figmodel} and Table 1. The red dashed line is the model that uses small grains; the wall has $a_{max}$=0.25 {\micron} and the optically thin region has $a_{min}$=0.005 {\micron} and $a_{max}$=0.25 {\micron}.  [See the electronic edition of the Journal for a color version of this figure.] 
}
\label{figvs}
\end{figure}

We detect [Ne II] fine-structure emission at 12.81 {\micron} $(^{2}P_{1/2} \rightarrow^{2}P_{3/2}$) in CS Cha (see Figure ~\ref{figvs}).  Using the IRAF tool $SPLOT$, we find the integrated line flux, $\int F_{\lambda} d\lambda$, to be 4.3 $\times$ 10$^{-14}$ ergs cm$^{-2}$ s$^{-1}$ or 1.3 $\times$ 10$^{29}$ ergs s$^{-1}$.

\citet[GNI07]{glassgold07} have recently suggested that 
circumstellar disks exposed to stellar X-rays would produce 
[Ne II] fine-structure emission at a level that would be detectable  
with $Spitzer$ in nearby star-forming regions.  
In the GNI07 model, the disk surface is both ionized and 
heated by stellar X-rays.  Ne ions (primarily Ne$^+$ and Ne$^{2+}$) 
are produced through X-ray ionization and destroyed by charge 
exchange with atomic hydrogen and radiative recombination. 
Alternatively, Hollenbach \& Gorti (in prep) propose that 
EUV photons from the stellar chromosphere and/or  
stellar accretion can create an ionized HII-region-like layer at the 
disk surface that can also produce significant [Ne II] emission.

\section{Discussion \& Conclusions}
\label{con}
CS Cha shows clear evidence of advanced dust evolution in relation to other transitional disks: it has significant grain growth and substantial settling in its outer disk.  
The dust in 
the wall of the 
outer disk in CS Cha has grown to much larger sizes (5 {\micron}) than the dust
in the transitional disks CoKu Tau$/$4, DM Tau, GM Aur, and TW Hya, all of which have maximum grain sizes in the outer wall of 0.25 {\micron} \citep{dalessio05, calvet05, calvet02, uchida04}, and CS Cha has large grains in the inner optically thin region as well, similar to TW Hya \citep{sargent06}.  
CS Cha also 
needs more settling in the outer disk ($\epsilon$ = 0.01) to fit the far-infrared and mm data, indicating much higher dust depletion in the upper layers
than found in GM Aur and DM Tau, which were modeled with $\epsilon$ = 0.1 (C05).  
In addition, CS Cha further stands out due to its large hole size, having the largest modeled to date at $\sim$ 43 AU.  This large inner hole serves to decrease the mid-infrared dust continuum and hence increase the line to continuum ratio, facilitating the detection of [Ne II] fine-structure emission.  

\begin{figure}
\epsscale{0.8}
\plotone{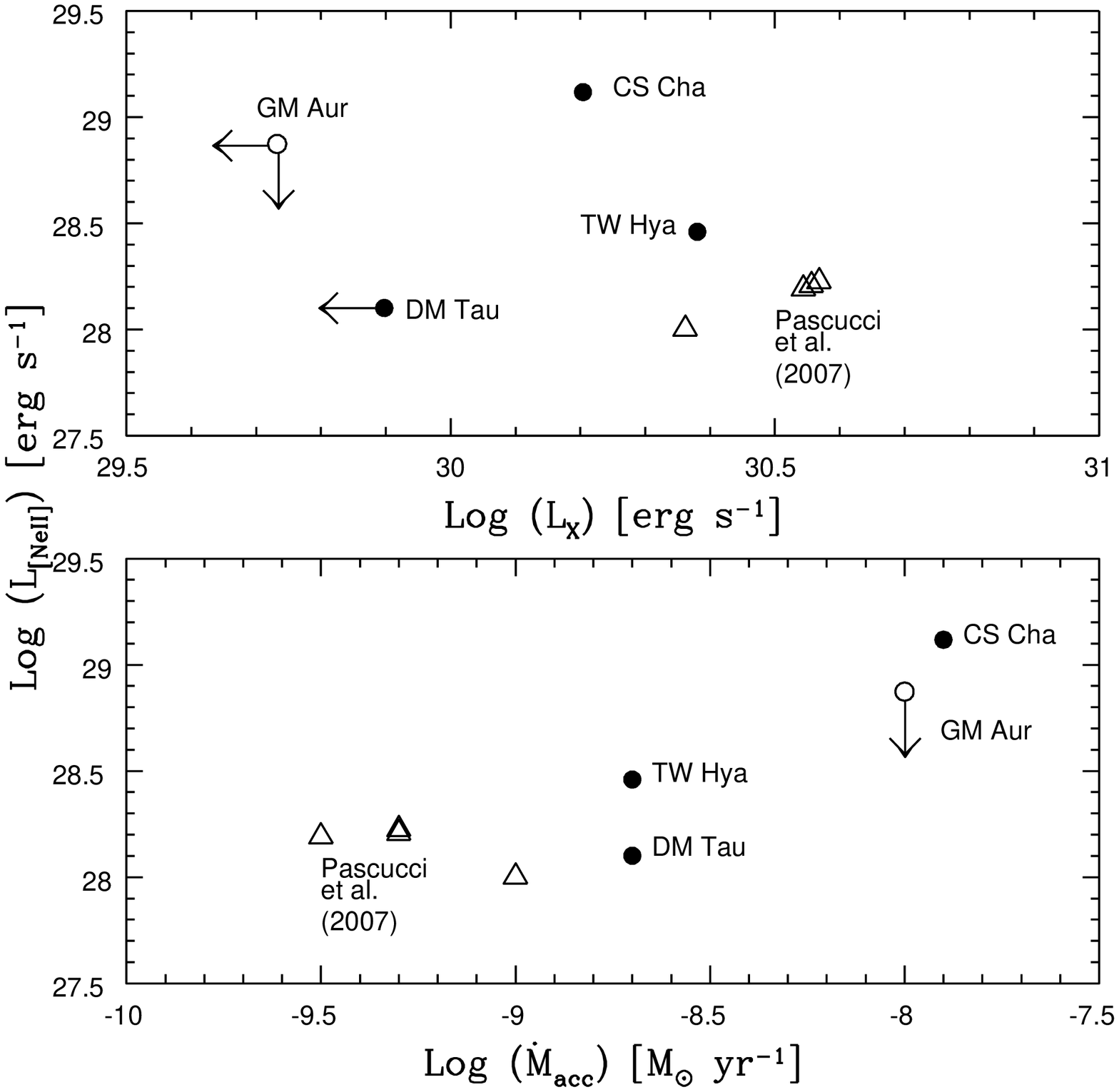}
\caption{Top: L$_{[Ne II]}$ vs. L$_{X}$.  There is no obvious correlation.  Triangles are from P07, circles correspond to [Ne II] luminosities derived in this work.  The L$_{X}$ for CS Cha and TW Hya are from \citet{feigelson93} and \citet{kastner99}, scaled to 160 and 55 pc, respectively.  We show upper limits for the L$_{[Ne II]}$ of GM Aur and the L$_{X}$ of GM Aur and DM Tau \citep{neuhauser95}.  All L$_{X}$ are from ROSAT and vary by a factor of $\sim$2.
Bottom: L$_{[Ne II]}$ vs. ${\mdot}$.  There is a possible positive correlation.  The ${\mdot}$ for CS Cha is derived here and ${\mdot}$ for GM Aur and DM Tau are from C05.  TW Hya's ${\mdot}$ is from veiling measurements but varies between 0.4 $\times$ 10$^{-9}$ to 10$^{-8}$ $\msun$ yr$^{-1}$ \citep{alencar02,muzerolle00}.}
\end{figure}

The luminosity of the [Ne II] 12.81 {\micron} line in CS Cha is strong compared to other [Ne II] detections in disks.  The luminosity seen in DM Tau is 1.3 $\times$ 10$^{28}$ ergs s$^{-1}$ and for GM Aur the 5 $\sigma$ upper limit is 7.4 $\times$ 10$^{28}$ ergs s$^{-1}$ (measured from spectra in C05).   TW Hya shows [Ne II] emission at 12.81 {\micron} with a luminosity of 3.9 $\times$ 10$^{28}$ ergs s$^{-1}$ (measured from Uchida et al.\ 2004).  In addition, the luminosities seen in four CTTS that were 
observed in the FEPS survey are $\sim$10$^{28}$ ergs s$^{-1}$ (P07). 

To gain insight into the origin of the
Neon emission we compare L$_{[Ne II]}$ to L$_{X}$ and ${\mdot}$
in Figure 4.  We see no apparent correlation between L$_{[Ne II]}$ and
L$_{X}$ as was reported by P07 and as would be expected from the X-ray heating model of GNI07,
which suggests that other processes like viscous heating, jets, and UV heating are involved (GNI07).  
We find a possible correlation between L$_{[Ne II]}$ and ${\mdot}$
which suggests that accretion related heating may play a substantial role in
producing [Ne II] emission.  
Accretion related processes that affect the disk heating are viscous
dissipation
and irradiation by UV emission from the accretion
shock region. Calvet et al. (2004) show that FUV fluxes scale with L$_{acc}$, and
a similar scaling is expected for the EUV; in turn, Hollenbach \& Gorti (in
prep.) show that EUV heating and ionization can result in significant [Ne II] emission.

Nevertheless, both the X-ray and EUV heating models
underpredict the L$_{[Ne II]}$ seen in CS Cha (GNI07; P07).  
We note
that the optically thin inner region of CS Cha is
populated by large grains, which would favor UV flux
penetration of the gas \citep{aikawa06} and possibly lead to enhanced
L$_{[Ne II]}$.  Further exploration of the gas and dust in other transitional disks is
necessary to test any link between these components.
Similarly, jets and outflows, which are
also characteristic of sources with higher accretion rates, may
contribute to a positive trend.  High resolution spectroscopy of CS Cha can explore
this contribution from the jet \citep{takami03}.   

CS Cha offers interesting insight into the dust and gas evolution of protoplanetary disks and encourages future studies.


\vskip -0.1in \acknowledgments{ We would like to thank David Hollenbach, Jon Miller, and an anonymous referee for
insightful comments.  This work is based on observations made with the
Spitzer Space Telescope.  NC and LH acknowledge support from NASA
Origins Grants NNG05GI26G and NNG06GJ32G. PD acknowledges grants from
CONACyT, M\`exico.}






\begin{thebibliography}
 
\bibitem[Alencar \& Batalha(2002)]{alencar02} Alencar, S.H.P. \& Batalha, C.  2002, \apj, 571, 378 
 
\bibitem[Alexander et al.(2006)]{alexander06} Alexander, R.D,  Clarke, C.J., \& Pringle, J.E. 2006, MNRAS, 369, 229

\bibitem[Aikawa \& Nomura(2006)]{aikawa06} Aikawa, Y., \& 
Nomura, H.\ 2006, \apj, 642, 1152 

\bibitem[Baraffe et al.(2002)]{baraffe02} Baraffe, I., et al. 2002, A\&A, 382, 563

\bibitem[Bergin et al.(2004)]{bergin04} Bergin, E., et al. 2004, ApJ, 614, L133 

\bibitem[Calvet et al.(1991)]{calvet91} Calvet, N., et al. 1991, ApJ, 380, 617

\bibitem[Calvet \& Gullbring(1998)]{calvetgullbring98} Calvet, N. \& Gullbring, E.  1998, \apj, 509, 802

\bibitem[Calvet et al.(2002)]{calvet02} Calvet, N., et al. 2002, ApJ, 568, 1008

\bibitem[Calvet et al.(2004)]{calvet04} Calvet, N., et al. 2004, AJ, 128, 1294



\bibitem[Calvet et al.(2005)]{calvet05} Calvet, N., et al., 2005, ApJL, 630, L185;C05 


\bibitem[D'Alessio et al.(1999)]{dalessio99}D'Alessio, P., et al.,1999,ApJ,527,893


\bibitem[D'Alessio et al.(2005)]{dalessio05} D'Alessio, P., et al., 2005, ApJ, 621, 461

\bibitem[D'Alessio et al.(2006)]{dalessio06} D'Alessio, P., et al., 2006, ApJ, 638, 314



\bibitem[Feigelson et al.(1993)]{feigelson93} Feigelson, E., et al. 1993, ApJ, 416, 623


\bibitem[Forrest et al.(2004)]{forrest04} Forrest, W.J., et al. 2004, ApJS, 154, 443

\bibitem[Furlan et al.(2006)]{furlan06} Furlan, E. et al., 2006, ApJS, 165, 568

\bibitem[Gauvin \& Strom(1992)]{gauvin92} Gauvin, I.S. \& Strom, K.M., 1992, ApJ, 385, 217


\bibitem[Glassgold, Najita, \& Igea(2004)]{glassgold04} Glassgold, A.E., Najita, J., \& Igea, J.2004, ApJ, 615, 972

\bibitem[Glassgold, Najita, \& Igea(2007)]{glassgold07} Glassgold, A.E., Najita, J., \& Igea, J. 2007, \apj, 656, 515; GNI07 


\bibitem[Gullbring et al.(1998)]{gullbring98} Gullbring, E., Hartmann, L., Brice\~{n}o, C., \& Calvet, N., 1998, ApJ, 492, 323

\bibitem[Hartigan(1993)]{hartigan93} Hartigan, P., 1993, AJ, 105, 1511

\bibitem[Henning et al.(1993)]{henning93} Henning et al, 1993, A\&A, 276, 129



\bibitem[Higdon et al.(2004)]{higdon04} Higdon, S.J.U, et al. 2004, PASP, 116, 975

\bibitem[Houck(2004)]{houck04}Houck, J. R., et al. 2004, ApJS, 154,18


\bibitem[Jayawardhana et al.(1999)]{jayawardhana99} Jayawardhana, R., et al., 1999, ApJ, 521, L129


\bibitem[Kastner, Huenemoerder, \& Schulz(1999)]{kastner99} Kastner, J.H., Huenemoerder, D.P., \& Schulz, N.S. 1999, \apj, 525, 837

\bibitem[Kenyon \& Hartmann(1995)]{KH95}Kenyon, S.~J.~\& Hartmann, L.\ 1995, \apjs, 101, 117


\bibitem[Lommen et al.(2007)]{lommen07} Lommen, D., 2007, A\&A, 462, 211

\bibitem[Luhman(2004)]{luhman04} Luhman, K.L., 2004, ApJ, 602, 816

\bibitem[Makovoz \& Marleau(2005)]{makovoz05}Makovoz and Marleau, 2005, PASP, 117, 1113

\bibitem[Mathis(1990)]{mathis90}Mathis, J.S. 1990, ARA\&A, 28, 37

\bibitem[Muzerolle et al.(2000)]{muzerolle00} Muzerolle, J., et al., 2000, ApJ, 535, L47


\bibitem[Najita et al.(2007)]{najita07} Najita, J.R. and Carr, J.S. and Glassgold, A.E. and Valenti, J.A. 2007, PPV

\bibitem[Neuhauser et al.(1995)]{neuhauser95} Neuhauser, R., et al.,  1995, A\&A, 297, 391


\bibitem[Pascucci et al.(2007)]{pascucci07} Pascucci et al, astro-ph/0703616; P07


\bibitem[Quillen et al.(2004)]{quillen04} Quillen, A.C., Blackman, E.G., Frank, A., \& Varniere, P. 2004,ApJ,612,L137


\bibitem[Rice et al.(2003)]{rice03} Rice et al., 2003, MNRAS, 342, 79




\bibitem[Sargent et al.(2006)]{sargent06}Sargent, B., et al. 2006, \apj, 645, 395



\bibitem[Su et al.(2006)]{su06}Su, K.Y.L., et al., 2006, ApJ, 653, 675

\bibitem[Takami et al.(2003)]{takami03}Takami,M., Dailey,J., \& Chrysostomou,A. 2003, A\&A, 397, 675

 

\bibitem[Uchida et al.(2004)]{uchida04} Uchida, K.I., et al., 2004, ApJS, 154, 439


\bibitem[Webb et al.(1999)]{webb99} Webb, R.A.  1999, ApJ, 512, 63


\bibitem[Werner et al.(2004)]{werner04}Werner, M. W., et al. 2004, ApJS, 154, 1

\bibitem[White \& Basri(2003)]{whitebasri03}White, R. \& Basri, G. 2003, \apj, 582, 1109

\bibitem[Whittet et al.(1997)]{whittet97}Whittet, D.C.B. et al. 1997, A\&A, 327, 1194








































 
 
 



\end{thebibliography}
\end{document}